\documentclass[11pt,twoside]{article}  % Leave intact
\usepackage{asp2006}
\usepackage{adassconf}

\begin{document}   % Leave intact

\paperID{P5.11}

\title{The Application of kd-tree in Astronomy}

\author{Dan \ Gao, Yanxia \ Zhang, Yongheng \ Zhao}
\affil{National Astronomical Observatories, Chinese Academy of
Sciences, China, 100012}

\contact{Dan Gao}
\email{zyx@lamost.org}

\paindex{Zhang, Y.}
\aindex{Zhao, Y.}
\aindex{Gao, D.}

\begin{abstract}          % Leave intact
The basic idea of the kd-tree algorithm is to recursively
partition a point set P by hyperplanes, and to store the obtained
partitioning in a binary tree. Due to its immense popularity, many
applications in astronomy have been implemented. The algorithm can
been used to solve a near neighbor problem for
cross-identification of huge catalogs and realize the
classification of astronomical objects. Since kd-tree can speed up
query and partition spaces, some approaches based on it have been
applied for photometric redshift measurement. We give the case
studies of kd-tree in astronomy to show its importance and
performance.
\end{abstract}

\section{Introduction of kd-tree}

K-dimensional tree (kd-tree), as a computer science term, is a
space-partitioning data structure for organizing points in a
$k$-dimensional space (Bentley, 1975). Technically, the letter $k$
refers to the numbers of dimensions. A 3-dimensional kd-tree can
be called as $3d$-tree. A kd-tree organizes datapoints in such a
way that once built, whenever a query arrives requesting a list
all points in a neighborhood, the query can be answered quickly
without needing to scan every single point. Each tree node
represents a subvolume of the parameter space, with the root node
containing the entire $k$ dimensional volume spanned by the data.
Non-leaf nodes have two children, obtained by splitting the widest
dimension of the parent's bounding box, the left child owning
those data points that are strictly less than the splitting value
in the splitting dimension, and the right child owning the
remainder of the parent's data points. A kd-tree is usually
constructed top-down, beginning with the full set of points and
then splitting in the center of the widest dimension. This
produces two child nodes, each with a distinct set of points.
Repeat this procedure recursively; a kd-tree can be constructed.
kd-trees have been successfully applied in astronomy for some
problems. Hsieh et~al. (2005) used the kd-tree algorithm to divide
up their sample to improve the redshift accuracy of galaxies. The
kd-tree method is used on the 5 flux-space indexing in the SDSS
Science Archive to partition the bulk data (Kunszt et~al. 2000).
In the following, we give three kinds of case studies with
kd-trees.

\section{Cross-matching of huge catalogs}

The cross-identification of catalogues from different bands,
especially from large survey projects, is the bottleneck of
multi-wavelength astronomical research, especially for data mining
and statistical study. Jim Gray (2005) present some performance
tests that using Zoning algorithm to facilitate large-scale query
and cross-match. Power et~al. (2004) introduce the plane sweep
technique to solve cross-matching problems with huge catalogues,
such as 2MASS, Tycho-2, USNOB1.0 and so on.

In some previous cross-match jobs, one object from catalogue A
needs to check with all the objects from catalogue B to obtain
matching counterparts, this requires a spatial join between tables
in databases. Since the computational complexity of these
cross-match algorithms are O($N^{2}$), major issues such as memory
limit and speed bottleneck are involved and make huge catalogues
cross-identification infeasible. So we introduce the Hierarchical
Triangular Mesh (HTM, http://skyserver.org/HTM/) spatial index
algorithm, and it gives a very efficient indexing method for
objects localized on the sphere. Moreover, kd-tree algorithm has
been involved to solve a near neighbor problem for huge catalogs
cross-identification.

Our solution is as follows: we firstly divide the sky into small
equal triangles by HTM, and then for each triangle we build a
kd-tree and find the nearest neighbor, and finally we judge
whether the nearest object is the cross-match counterpart. Assumed
that there are two tables: table A and table B, and the record
number in table A is smaller than that in table B. We map each
object (ra, dec) to a HTM index number for each catalogue table
and divide the sky into equal triangles. For each triangle, we
insert objects from table B into a kd-tree and build it, and then
for each object from table A we find the nearest neighbor or
nearest N neighbors in the kd-tree. Finally, we can determine
whether the neighbor object is the association by analyzing the
distribution of distances between the object from table A and the
neighbor object from table B, and corresponds to 3$\sigma$ cutoff
(Zhang, 2003).

Table 1 listed some applications of huge catalog cross-match using
the above method. For example, the matching of SDSS spectroscopic
quasar sample with 2MASS catalogue takes 5,033 seconds. The result
indicated that cross-matching two catalogues consisting of half a
million or ten million objects can be achieved in around one hour.
When one of the catalogues consists of a billion or ten billion
objects the total time taken is about 10 hours or one day.

\begin{table}
\caption{Applications of huge catalogs cross-matching.}
\begin{center}
\begin{tabular}{lllllll}
\tableline \tableline
Catalog A&records&disk &Catalog B &records &disk & time \\
& & (M)& & &(G) & (sec)\\
 \tableline
       RASS-FSC   & 105,924 & 18 & Tycho-2 & 2,539,913 &  0.4  &  3567 \\
       SDSS quasars & 76,989 & 56 & 2MASS  & 470,992,970 & 123 &  5033 \\
       GSPC2.4 & 554,007 & 65 & USNO-B1.0 &  1,045,096,352  & 172  &  85720 \\

\tableline
\end{tabular}
\end{center}
\end{table}

\section{Classification}

We collected photometric data of quasars and stars with spectra
measurement from SDSS DR5. After crossing out the missing data, we
obtained the sample contains 76,949 quasars and 108,744 stars.
Train-test method is used to divide the sample into two parts:
124,415 for training a classifier and 61,281 for testing the
classifier to get the classification rate. We use two different
magnitudes: model magnitudes and model magnitudes with reddening
correction (hereafter short for dereddened magnitudes) from SDSS
data. The four model color index ($u-g$, $g-r$, $r-i$, $i-z$) and
the model $r$ magnitude are taken as the first set of input
parameters for kd-tree, and then the four dereddened colors and
the dereddened $r$ magnitude are as input patterns.

The results are given in Table 2. When considering model
magnitudes, the accuracy of quasars and stars are 96.41\% and
97.87\%, respectively. The accuracy is 96.37\% for quasars and
97.76\% for stars given dereddened magnitudes. For the two input
patterns, the total accuracy is 97.26\% and 97.19\%, respectively.
Therefore, we summarized that the performance of the input pattern
based on model magnitudes adds up to a higher accuracy than that
based on dereddened magnitudes. For any input pattern, the
accuracy is rather high, more than 97\%, and the running time is
not more than 1 minute. In Table 2, we concluded that a kd-tree
shows its superiority to separate quasars from stars in respect of
both accuracy and speed. As a result, the classifiers trained with
kd-tree can be used to classify the unclassified sources and be
applicable to preselect quasar candidates for large surveys, such
as the Chinese Large Sky Area Multi-Object Fiber Spectroscopic
Telescope (LAMOST).

\begin{table} \caption{Separate quasars from stars using kd-tree.}
\begin{center}
\begin{tabular}{r|ll|ll}
\tableline \tableline
Sample&input &pattern&dereddened &input\ \ \  pattern \\
\tableline
classified$\downarrow$known$\to$& quasars &stars& quasars &stars\\
\tableline
       quasars   &  24483  &  766 & 24474 & 803 \\
       stars &  911   &35121 & 921 & 35085 \\
\tableline
       Accuracy & 96.41\% & 97.87\% & 96.37\% & 97.76\% \\
\tableline
       Total accuracy & 97.26\% && 97.19\% &\\
\tableline
\end{tabular}
\end{center}
\end{table}

\section{Photometric Redshifts}

We collected photometric data of galaxies with spectra measurement
from SDSS DR5. The outlying data and the missing data are crossed
out. In order to estimate photometric redshifts of galaxies
derived from SDSS DR5, we obtained all objects satisfying the
following criteria (Vanzella et al. 2004): (1) $r$-band Petrosian
magnitude $r <$ 17.77; (2) the spectroscopic redshift confidence
must be greater than 0.95 and there must be no warning flags. This
led to 375,929 galaxies, which are randomly partitioned into
training set (251,872) and test set (124,057). In this experiment,
we directly used the four dereddened colors as input pattern, and
then we adopted the four dereddened colors and the dereddened $r$
magnitude as the input of kd-tree.

The dispersion ($\sigma_{rms}$) with each set of parameters to
estimate photometric redshifts are listed in Table 3. When taking
$u-g$, $g-r$, $r-i$, $i-z$ as inputs, the rms scatter in testing
set is 0.0212. When taking $u-g$, $g-r$, $r-i$, $i-z$ and $r$ as
inputs, the rms error is up to 0.0232. Table 3 shows that the
performance based on the combination of four colors and $r$
magnitude is superior to that based on only four colors.
\begin{table}
\caption{Photometric redshift measurement with kd-tree.}
\begin{center}
\begin{tabular}{ll}
\tableline \tableline Input Parameters  & $\sigma_{rms}$ \\
\tableline  $dereddened (u-g,g-r,r-i,i-z)$ & 0.0212 \\
            $dereddened (u-g,g-r,r-i,i-z,r)$ & 0.0232 \\
  \tableline
\end{tabular}
\end{center}

\end{table}

\section{Conclusion}

From the case studies, it is concluded that kd-trees have a wide
application in astronomy due to its own characteristics. Kd-trees
are useful data structure for many applications, such as searches
involving a multi-dimensional search key. Therefore many
researches related to optional and fast search may be searched
with the kd-tree approach. With the increase of astronomical data
in volume, kd-tree may play a more important role.

\acknowledgments This paper is funded by National Natural Science
Foundation of China under grants No.10473013, No.10778724 and No.90412016.

\end{document}